\newcommand{\diracslash}[1]{#1\llap{/\kern2pt}}

\newcommand{\be}{\begin{equation}}
\newcommand{\ee}{\end{equation}}
\newcommand{\bea}{\begin{eqnarray}}
\newcommand{\eea}{\end{eqnarray}}
\newcommand{\ba}[1]{\begin{array}{#1}}
\newcommand{\ea}{\end{array}}

\newcommand{\bt}{\begin{tabular}}
\newcommand{\et}{\end{tabular}}

\newcommand{\beas}{\begin{eqnarray*}}
\newcommand{\eeas}{\end{eqnarray*}}

\documentclass[preprint,prd,aps,floats,nofootinbib,floatfix]{revtex4}
\usepackage{graphicx}
\begin{document}

\title{Light vector mesons ($\omega$, $\rho$ and $\phi$)
in strong magnetic\\
fields: A QCD sum rule approach}
\author{Amruta Mishra}
\email{amruta@physics.iitd.ac.in}
\affiliation{Department of Physics, Indian Institute of Technology, Delhi,
Hauz Khas, New Delhi -- 110 016, India}

\author{Ankit Kumar}
\email{ankitchahal17795@gmail.com}
\affiliation{Department of Physics, Indian Institute of Technology, Delhi,
Hauz Khas, New Delhi -- 110 016, India}

\author{Pallabi Parui}
\email{pallabiparui123@gmail.com}
\affiliation{Department of Physics, Indian Institute of Technology, Delhi,
Hauz Khas, New Delhi -- 110 016, India}

\author{Sourodeep De}
\email{sourodeepde2015@gmail.com}
\affiliation{Department of Physics, Indian Institute of Technology, Delhi,
Hauz Khas, New Delhi -- 110 016, India}

\begin{abstract}
The mass modifications of the light vector mesons 
($\omega$, $\rho$ and $\phi$)
are investigated in asymmetric nuclear matter in the presence
of strong magnetic fields, using a QCD sum rule approach.
These are computed from the medium modifications
of the non-strange and strange light quark condensates
as well as scalar gluon condensate.
The quark and gluon condensates
are calculated from the medium changes of the scalar fields
(non-strange and strange) and a scalar dilaton field 
in the magnetized nuclear matter, within a chiral SU(3) model.
The scalar dilaton field within the model breaks the scale
invariance of QCD and simulates the
gluon condensate. The anomalous magnetic
moments for the nucleons are taken into account in the present 
study. 

\end{abstract}
\maketitle

\def\bfm#1{\mbox{\boldmath $#1$}}

\section{Introduction}
The study of properties of hadrons under extreme conditions, e.g,
high densities and/or temperatures, is an important topic
of contemporary research in strong interaction physics.
The subject is of relevance to ultrarelativistic heavy ion 
collision experiments, where the experimental observables of these
high energy nuclear collisions are affected by
the medium modifications of the hadrons.
Furthermore, the heavy colliding nuclei have large isospin asymmetry
as the number of the neutrons is much larger than the number
of protons of these nuclei. It is thus important to study the
effects of isospin asymmetry on the hadron properties.
The estimation of huge magnetic fields being created in non-central 
ultra-relativistic heavy ion collision experiments necessitates
the study of magnetic field effects on the properties of the
hadrons. The medium modifications of the hadrons have been 
studied extensively in the literature. The different formalisms
for these studies are the effective hadronic 
models, e.g., Quantum Hadrodynamics  (QHD) model \cite{walecka}, 
the QCD sum rule (QCDSR) approach \cite{svznpb1,svznpb2}, 
the quark meson coupling (QMC) model\cite{qmc},  
the chiral effective models, as well as using the coupled channel approach.
The models like Nambu Jona Lasinio model, which simulate
the spontaneous chiral symmetry breaking of QCD
(through four fermion interactions), 
 have been extensively used in the literature
\cite{klevansky, buballanjlphysrep, hatkunihironjl, njlhmjcp,
njlamhm}, 
to study the strongly interacting matter.
The AdS/CFT correspondence and the conjecture of gravity/gauge duality
\cite{adscft} have also been used to study the hadrons \cite{hadadsqcd}.

The in-medium masses of the light vector mesons ($\rho$, $\omega$ and $\phi$)
are studied in the present work using a QCD sum rule approach
\cite{hatlee,hatlee2,hatnpb,asakawako,zschocke,klinglnpa,klinglzpa,
kwonprc2008,kwonweiseprc2010,adami,thomas4quark,Abhee}, 
in asymmetric nuclear matter in the presence of strong magnetic fields.
The medium modifications are due to the changes of the 
light quark condensates and gluon condensate in the magnetized
isospin asymmetric hadronic matter, which are calculated 
within mean field approximation,
from the changes in the expectation values of non-strange ($\sigma$) and
strange ($\zeta$) scalar-isoscalar fields, the third component
of a scalar isovector field ($\delta$) and a dilaton field
($\chi$) from their vacuum values, 
within a chiral SU(3) model \cite{kristof1,papa}. The quark condensates, 
are obtained from the explicit symmetry breaking term 
within the chiral SU(3) model, in terms of the scalar fields,
$\sigma$, $\zeta$ and $\delta$, and the gluon condensate
is related to the dilaton field, $\chi$, which mimics
the scale symmetry of QCD, through a logarithmic potential.
The model has been used to describe nuclear matter \cite{kristof1}, 
finite nuclei \cite{papa} 
and the bulk properties of (proto) neutron stars \cite{nstar}.
Using the chiral SU(3) model, 
the vector mesons have also been studied \cite{vecm}, 
accounting for the Dirac polarization effects 
\cite{vecmwalecka}.
The model has been used to study the kaons and antikaons
in isospin asymmetric nuclear (hyperonic) matter 
\cite{kaon_antikaon,isoamss,isoamss1,isoamss2}.
The model has been generalized to the charm and bottom sectors and
the in-medium masses of open charm 
\cite{amdmeson,amarindamprc,amarvdmesonTprc,amarvepja,DP_AM_Ds},
open bottom mesons \cite{DP_AM_bbar,DP_AM_Bs},
charmonium \cite{amarvepja,jpsi_etac_AMarv} and 
bottomonium states \cite{AM_DP_upsilon}.
Using the mass modifications of the charmonium states
and the open charm mesons, the partial decay widths of the
charmonium states to $D\bar D$ pair
in the hadronic medium have been studied 
using $^3P_0$ model \cite{3p0,friman,amarvepja}. 
The in-medium partial
decay widths of the charmonium (bottomonium)
to $D\bar D$ ($B\bar B$) \cite{amspmwg,amspm_upsilon}
have  also been studied using a field theoretic model 
for composite hadrons, from the mass modifications
of these heavy flavour mesons calculated within the
chiral effective model.
The effects of magnetic fields on these heavy flavour mesons
($D$, $B$, charmonium and bottomonium states)
in asymmetric nuclear matter have also been studied 
\cite{dmeson_mag,bmeson_mag,charmonium_mag,upsilon_mag,jpsi_etac_mag}, 
by including the coupling terms with the electromagnetic field
to the baryons in the Lagrangian density of the chiral
effective model. The masses of these heavy flavour mesons
have been studied accounting for 
the effects of the anomalous magnetic moments of the nucleons.
The light vector mesons ($\omega$, $\rho$ and $\phi$) 
in strange hadronic matter have been 
studied using a QCD sum rule approach \cite{am_vecqsr}, 
using the medium dependent quark and gluon condensates,
calculated within the chiral SU(3) model.
In the present investigation, we compute the
medium modifications of these vector mesons
in asymmetric nuclear matter in the presence of
strong magnetic fields using QCD sum rule approach,
with the quark and gluon condensates obtained
from the scalar fields and dilaton field
within the chiral model.


The outline of the paper is as follows : In section II, we describe
briefly the chiral $SU(3)$ model used to calculate the quark and
gluon condensates in the nuclear medium in the presence of strong
magnetic fields. The in-medium values of these condensates
are calculated from the medium changes of the scalar fields
of the explicit symmetry breaking term and of the dilaton field,
which mimics the gluon condensate of QCD in the chiral SU(3) model.
In section III, we present the QCD sum rule approach 
using which the in-medium masses of the light vector mesons
($\omega$, $\rho$, $\phi$)
are studied. Section IV discusses the results of the mass modifications
of the light vector mesons ($\omega$, $\rho$ and $\phi$)
in the magnetized  asymmetric nuclear matter.
In section V, we summarize the findings of the present investigation.

\section{The hadronic chiral $SU(3) \times SU(3)$ model }
We use an effective chiral $SU(3)$ model \cite{papa} to obtain the in-medium
quark and gluon condensates for the study of modifications
of the masses of the light vector mesons using a QCD sum rule
approach. The model is based on the nonlinear realization of chiral 
symmetry \cite{weinberg, coleman, bardeen} and broken scale invariance 
\cite{papa,kristof1,vecm}. 
The concept of broken scale invariance leading to the trace anomaly 
in QCD, $\theta_{\mu}^{\mu} = \frac{\beta_{QCD}}{2g} 
{G^a}_{\mu\nu} G^{\mu\nu a}$, where $G_{\mu\nu}^{a} $ is the 
gluon field strength tensor of QCD, is simulated in the effective 
Lagrangian at tree level through the introduction of 
the scale breaking terms \cite{sche1,ellis}
\begin{equation}
{\cal L}_{scalebreaking} =  -\frac{1}{4} \chi^{4} {\rm {ln}}
\Bigg ( \frac{\chi^{4}} {\chi_{0}^{4}} \Bigg ) + \frac{d}{3}{\chi ^4} 
{\rm {ln}} \Bigg ( \bigg ( \frac { \sigma^{2} \zeta }{\sigma_{0}^{2} 
\zeta_{0}}\bigg) \bigg (\frac {\chi}{\chi_0}\bigg)^3 \Bigg ).
\label{scalebreak}
\end{equation}
The Lagrangian density corresponding to the dilaton field, $\chi$
leads to the trace of the energy momentum tensor as 
\cite{heide1,jpsi_etac_AMarv,am_vecqsr}
\begin{equation}
\theta_{\mu}^{\mu} = \chi \frac{\partial {\cal L}}{\partial \chi} 
- 4{\cal L} 
= -(1-d)\chi^{4}.
\label{tensor1}
\end{equation}
Equating the trace of the energy momentum tensor arising
from the trace anomaly of QCD with that of the present chiral model
given by equation (\ref{tensor1}),
gives the relation of the dilaton field to the scalar gluon condensate.
The trace of the energy momentum tensor in QCD is given as
\cite{cohen},
\begin{equation}
T_{\mu}^{\mu} = \sum _{q_i =u,d,s } m_{q_i} \bar {q_i} {q_i} + 
\langle \frac{\beta_{QCD}}{2g} 
G_{\mu\nu}^{a} G^{\mu\nu a} \rangle  \equiv  -(1 - d)\chi^{4}. 
\label{tensor2m}
\end{equation}
In the above, the first term of the energy-momentum tensor, 
within the chiral SU(3) model is the negative of the explicit
chiral symmetry breaking term. This relates the light quark condensates 
to the values of the scalar fields $\sigma$, $\zeta$ and $\delta$ 
in the mean field approximation as \cite{am_vecqsr}
\begin{eqnarray}
m_u\langle \bar u u \rangle 
= \frac{1}{2}m_{\pi}^{2} f_{\pi} (\sigma+\delta),\nonumber\\
m_d \langle \bar d d \rangle
= \frac{1}{2}m_{\pi}^{2} f_{\pi} (\sigma-\delta),\nonumber \\
m_s\langle \bar s s \rangle 
= \Big( \sqrt {2} m_{k}^{2}f_{k} - \frac {1}{\sqrt {2}} 
m_{\pi}^{2} f_{\pi} \Big) \zeta.
\label{qqbar}
\end{eqnarray}

Using the QCD $\beta$ function occurring in the right hand side 
of equation (\ref{tensor2m}) at one loop order,
for $N_{c}=3$ colors and $N_{f}=3$ flavors,
one gets the dilaton field related to the scalar
gluon condensate as
\begin{equation}
\left\langle  \frac{\alpha_{s}}{\pi} {G^a}_{\mu\nu} {G^a}^{\mu\nu} 
\right\rangle =  \frac{8}{9} \Bigg [(1 - d) \chi^{4}
+ \left( m_{\pi}^{2} f_{\pi} \sigma
+ \Big( \sqrt {2} m_{k}^{2}f_{k} - \frac {1}{\sqrt {2}} 
m_{\pi}^{2} f_{\pi} \Big) \zeta \right) \Bigg ]. 
\label{chiglu}
\end{equation}
The coupled equations of motion for the non-strange scalar isoscalar 
field $\sigma$, scalar isovector field, $\delta$, the
strange scalar field $ \zeta$, and the dilaton 
field $\chi$, derived from the Lagrangian density
of the chiral SU(3) model,
are solved to obtain the values of these fields
in the asymmetric nuclear medium in the presence of 
magnetic field. 

\section{QCD sum rule approach}
In the present section, we briefly describe the QCD sum rule approach
used to study the properties of the light vector mesons 
($\omega$, $\rho$, $\phi$) in the nuclear medium in presence of
a magnetic field. The study uses the values of quark and gluon
condensates in the magnetized nuclear medium obtained in a 
chiral SU(3) model as described in the previous section.
The current current correlation function for the
vector meson, V(=$\omega$,$\rho$, $\phi$) is written as

\begin{equation}
\Pi^V_{\mu \nu}(q)= i\int d^4x e^{iq \cdot x} 
\langle 0| T j^V_\mu (x) j^V _\nu (0)|0\rangle,
\end{equation}
where $T$ is the time ordered product and $J^V_\mu$ is the 
current for the vector meson, $V=\rho,\omega,\phi$.
Current conservation gives the structure of the correlation function
as
\begin{equation}
\Pi^V_{\mu \nu}(q)=\left (g_{\mu \nu}-\frac{q_\mu q_\nu}{q^2}
\right) \Pi^V (q^2).
\end{equation}
In large space-like region, $Q^2=-q^2 >> $ 1 GeV$^2$, the scalar 
correlation function $\Pi^V (q^2)$ 
for the light vector mesons ($\omega$, $\rho$
and $\phi$) can be written in terms of the operator product
expansion (OPE) as \cite{klinglnpa,kwonprc2008}
\begin{equation}
12\pi^2{\tilde \Pi^V} (q^2=-Q^2)=
d_V \Big [ -c^V_0 \ln \Big (\frac { Q^2}{\mu^2}\Big )
+\frac {c^V_1}{Q^2} + \frac {c^V_2}{Q^4} +\frac {c^V_3}{Q^6}+\cdots \Big ]
\label{qcdope}
\end{equation}
where, $\tilde \Pi^V (q^2=-Q^2)=\frac{ \Pi^V (q^2=-Q^2)}{Q^2}$,
$d_V$=3/2, 1/6 and 1/3 for $\rho$, $\omega$ and $\phi$ mesons
respectively, and,
$\mu$ is a scale chosen to be 1 GeV in the present
investigation \cite{am_vecqsr}. 
The leading term in the OPE, given by the first term,
is calculated in the perturbative QCD.
The coefficients $c^V_i (i=1,2,3)$ in the OPE
contain the informations of the nonperturbative effects
of QCD in terms of the quark and gluon condensates, as well as,
of the Wilson coefficients \cite{hatnpb,zschocke}.
The Wilson coefficients are taken as medium independent, with all the
medium effects incorporated into the quark and gluon condensates
\cite{hatnpb,zschocke,klinglnpa,kwonprc2008,kwonweiseprc2010}.
After Borel transformation, the correlator for the vector meson
can be written as
\begin{equation}
 12 \pi^2 \tilde  \Pi^V (M^2)=d_V \Big [ c^V_0 M^2 +c^V_1+ \frac{c^V_2}{M^2}
+\frac{ c^V_3}{2M^4}\Big ]
\label{corropeborel}
\end{equation} 
On the phenomenological side 
the correlator function, $\tilde \Pi^V (Q^2)$ 
can be written as
\begin{equation}
12 \pi^2 \tilde \Pi^V _{phen}(Q^2)
=\int _0 ^\infty ds \frac{R^V_{phen}(s)}{s+Q^2}
\label{corrphen}
\end{equation}
where $R^V_{phen}(s)$ is the spectral density proportional to the
imaginary part of the correlator
\begin{equation}
R^V_{phen}(s)={12 \pi} {\rm {Im}} \Pi^V _{phen} (s).
\end{equation}
On Borel transformation, equation (\ref{corrphen}) reduces to 
\begin{equation}
12 \pi^2 \tilde \Pi^V (M^2)=\int _0 ^\infty d s e^{-s/{M^2} }
R^V_{phen}(s)
\label{corrphenborel}
\end{equation}
Equating the correlation functions from the phenomenological side
given by equation (\ref{corrphenborel}) to that from the operator
product expansion given by equation (\ref{corropeborel}),
we obtain,
\begin{equation}
\int _0 ^\infty d s e^{-{s}/{M^2} }
R^ V_{phen} (s)+12 \pi^2 \Pi^ V(0) ={d_V}
\Big [ c^V_0 M^2 +c^V_1+ \frac{c^V_2}{M^2}
+\frac{ c^V_3}{2M^4}\Big ],
\label{qsr}
\end{equation}
where the second term in the left hand side of the above
equation is the contribution due to scattering of the
vector meson with the baryons in the hadronic medium.
In the nuclear medium as is the case of the present 
work of the study of in-medium masses of vector meson,
$\Pi^ V(0)=\frac {\rho_B}{4M_N}$ 
for V=$\omega$,$\rho$. Further, $\Pi^V(0)$ vanishes
for $\phi$ meson, since the $\phi$-meson nucleon coupling
is zero in the parametrization 
for the vector meson--baryon interactions within 
the chiral SU(3) model \cite{papa}. This is a reasonable 
assumption due to the fact that the $\phi$-meson nucleon 
scattering amplitude
is negligibly small as compared to the $\omega$-nucleon
as well as $\rho$-nucleon scattering amplitudes 
\cite{klinglnpa,hatlee2,bochkarev,florkowski}.
The spectral density is assumed to be of the form
of a resonance part ${R^V}_{phen}^{(res)}(s)$ 
and a perturbative continuum as \cite{klinglnpa,am_vecqsr}
\begin{equation}
R^V_{phen}(s) ={R^V}_{phen}^{(res)}(s) \theta (s^V_0-s)
+{d_V} c^V_0 \theta (s-s^V_0)
\label{qsr1}
\end{equation}
For $M > \sqrt {s^V_0}$, the exponential function in the integral of
the left hand side of the equation (\ref{qsr}) is expanded 
in powers of $s/M^2$ for $s < s^V_0$ and
one obtains the Finite energy sum rules (FESR) \cite{klinglnpa}
by equating the powers in $1/{M^2}$ of both sides of 
equation (\ref{qsr}). 
In the hadronic medium, the FESRs are obtained as 
\cite{am_vecqsr}
\begin{equation}
F^*_V =
{d_V} ({c^V_0} {{s^*}^V_0} +{c^V_1}) -12\pi^2 \Pi^V(0) 
\label{fesr1mf}
\end{equation}
\begin{equation}
F^*_V {m^*_V}^2=
{d_V} \Big (
\frac {({s^*}^V_0)^2 c^V_0}{2}-{c^*}_2^V \Big )
\label{fesr2mf}
\end{equation}
\begin{equation}
F^*_V {m^*_V}^4=
{d_V} \Big (
\frac{({s^*}^V_0)^3}{3} c^V_0 +{c^*}_3^V \Big )
\label{fesr3mf}
\end{equation}
The coefficient ${c^*}^V_2$ contains the quark and gluon
condensates in the medium, and ${c^*}^V_3$ corresponds 
to the four quark condensate, which is calculated using
a factorization method \cite{svznpb2} 
along with a parameter $\kappa_i$,$i=u,d,s$ which measures the
deviation from exact factorization ($\kappa_i$=1).
For the nonstrange vector mesons, $\rho$ and $\omega$,
these coefficients are given as
\begin{equation}
c_0 ^{(\rho,\omega)}=1+\frac{\alpha_s (Q^2)}{\pi},\;\;\;\;
c_1 ^{(\rho,\omega)}=-3 (m_u ^2 +m_d ^2)
\label{c0c1rhomg}
\end{equation}
\begin{equation}
{c^*}_2 ^{(\rho,\omega)}= \frac {\pi^2}{3}
\langle \frac {\alpha_s}{\pi} G^{\mu \nu} G_{\mu \nu}
\rangle + 4\pi^2 \langle m_u \bar u u +m_d \bar d d \rangle
\label{c2rhomg}
\end{equation}
\begin{equation}
{{c^*}_3}^{(\rho,\omega)}=
-\alpha_s \pi^3\times \frac{448}{81} \kappa_q (\langle \bar u u \rangle^2
+ \langle \bar d d \rangle^2),
\label{c3rhomgf}
\end{equation}
where, we take $\kappa_u \simeq \kappa_d =\kappa_q$. 
For $\phi$ meson, these coefficients are given as \cite{klinglnpa,svznpb1}
\begin{equation}
c_0 ^{\phi}=1+\frac{\alpha_s (Q^2)}{\pi},\;\;\;\;
c_1 ^{\phi}=-6 {m_s}^2 
\label{c0c1phi}
\end{equation}
\begin{equation}
{c^*}_2 ^{\phi}= \frac {\pi^2}{3}
\langle \frac {\alpha_s}{\pi} G^{\mu \nu} G_{\mu \nu}
\rangle + 8\pi^2 \langle m_s \bar s s \rangle
\label{c2phi}
\end{equation}
\begin{eqnarray}
{{c^*}_3}^{\phi} 
&=& -8\pi^3 \times \frac{224}{81} \alpha_s \kappa_s 
\langle \bar s s \rangle ^2.
\label{c3phif}
\end{eqnarray}
Solving the FESR for vector meson, V ($\rho$, $\omega$, $\phi$)
in vacuum, assuming the vacuum mass of the vector meson,
determines the value of the coefficient $\kappa_q$ ($\kappa_s$),
of the 4-quark condensate, along with the parameters,
$F_V$ and $s_0^V$ in vacuum \cite{am_vecqsr}.
The equations (\ref{fesr1mf}), (\ref{fesr2mf}) and (\ref{fesr3mf})
are then solved to obtain the medium dependent mass,
$m^*_V$, the scale ${s^*}_0^V$ and $F^*_V$ for the vector meson,
$V$, using the value of $\kappa_i$ as determined from 
the FESR in vacuum.

\begin{table}
\begin{tabular}{|c|c|c|c|c|c|}
\hline
\multicolumn{2}{|c|}{
} & \multicolumn{2}{|c|}{$ \eta=0 $} 
&  \multicolumn{2}{|c|}{$ \eta=0.5 $}\\
\cline{3-6}
\multicolumn{2}{|c|}
 {${eB}/{m_\pi^2}$}
 &  $  \rho_B=\rho_0 $ & $ \rho_B=2\rho_0 $ & 
$ \rho_B=\rho_0 $ & $ \rho_B=2\rho_0 $ \\
\hline
\multicolumn{2}{|c|}{
{0}
}   & 
{777}  & 953.7   & 
{788.8}  & 
{965.6}  \\
\hline
{
4}& (a) 
&  773.4 & 950 & 
787.24  & 964.2  \\
\cline{2-6}
& (b) 
 & 775.7 & 953.4 & 
792.5 & 967.56        \\
\hline
{
8
}&  (a)
& 772.8 & 947.8 & 
787.24 & 964.2 \\
\cline{2-6}
& (b)
& 775.6 & 952.4 & 
793.74 & 972 \\
\hline
{
12
}&  (a)
& 772.34 & 947.3 & 
787.24 &964.2\\
\cline{2-6}
& (b)
&775.87 & 953.34 & 
795.3 & 974.1\\
\hline
\end{tabular}
\vskip 0.1in
\caption{In-medium masses for $\omega$ meson in magnetized nuclear
matter for densities of $\rho_0$ and 2$\rho_0$, asymmetric parameter,
$\eta$=0 and 0.5 and for magnetic fields, $eB/{m_\pi^2}$ as 4,8 and
12. These masses are compared with the in-medium masses
of $\omega$ meson for zero magnetic field.}
\label{table1}
\end{table}

\begin{table}
\begin{tabular}{|c|c|c|c|c|c|}
\hline
\multicolumn{2}{|c|}{
} & \multicolumn{2}{|c|}{$ \eta=0 $} 
&  \multicolumn{2}{|c|}{$ \eta=0.5 $}\\
\cline{3-6}
\multicolumn{2}{|c|}
 {${eB}/{m_\pi^2}$}
 &  $  \rho_B=\rho_0 $ & $ \rho_B=4\rho_0 $ & 
$ \rho_B=\rho_0 $ & $ \rho_B=4\rho_0 $ \\
\hline
\multicolumn{2}{|c|}{
{0}
}   & 
{622.2}  & 391   & 
{636.3}  & 
{473.3}  \\
\hline
{
4}& (a) 
&  618 & 374.9 & 
634.5  & 468.5  \\
\cline{2-6}
& (b) 
 & 620.7 & 398 & 
640.85 & 477.5        \\
\hline
{
8
}&  (a)
& 617.26 & 335.8 & 
634.5 & 468.5 \\
\cline{2-6}
& (b)
& 620.65 & 411.7 & 
642.3 & 502.8 \\
\hline
{
12
}&  (a)
& 616.68 & 315.8 & 
634.5 &468.5\\
\cline{2-6}
& (b)
&620.9 & 433.7 & 
644.2 & 527.6\\
\hline
\end{tabular}
\vskip 0.1in
\caption{In-medium masses for $\rho$ meson in magnetized nuclear
matter for densities of $\rho_0$ and 4$\rho_0$, asymmetric parameter,
$\eta$=0 and 0.5 and for magnetic fields, $eB/{m_\pi^2}$ as 4,8 and
12. These masses are compared with the in-medium masses
of $\rho$ meson for zero magnetic field.}
\label{table2}
\end{table}

\begin{table}
\begin{tabular}{|c|c|c|c|c|c|}
\hline
\multicolumn{2}{|c|}{
} & \multicolumn{2}{|c|}{$ \eta=0 $} 
&  \multicolumn{2}{|c|}{$ \eta=0.5 $}\\
\cline{3-6}
\multicolumn{2}{|c|}
 {${eB}/{m_\pi^2}$}
 &  $  \rho_B=\rho_0 $ & $ \rho_B=4\rho_0 $ & 
$ \rho_B=\rho_0 $ & $ \rho_B=4\rho_0 $ \\
\hline
\multicolumn{2}{|c|}{
{0}
}   & 
{1001.49}  & 998.79   & 
{1001.79}  & 
{998.38}  \\
\hline
{
4}& (a) 
&  1001.11 & 998.4 & 
1001.6  & 997.85  \\
\cline{2-6}
& (b) 
 & 1001.28 & 998.08 & 
1002.03 & 997.74        \\
\hline
{
8
}&  (a)
& 1000.9 & 998.9 & 
1001.6 & 997.85 \\
\cline{2-6}
& (b)
& 1001.28 & 997.87 & 
1002.14 & 997.5 \\
\hline
{
12
}&  (a)
& 1001.09 & 999.2 & 
1001.6 &997.85\\
\cline{2-6}
& (b)
&1001.22 & 997.63 & 
1002.28 & 997.38\\
\hline
\end{tabular}
\vskip 0.1in
\caption{In-medium masses for $\phi$ meson in magnetized nuclear
matter for densities of $\rho_0$ and 4$\rho_0$, asymmetric parameter,
$\eta$=0 and 0.5 and for magnetic fields, $eB/{m_\pi^2}$ as 4,8 and
12. These masses are compared with the in-medium masses
of $\phi$ meson for zero magnetic field.}
\label{table3}
\end{table}

\begin{figure}
\includegraphics[width=16cm,height=16cm]{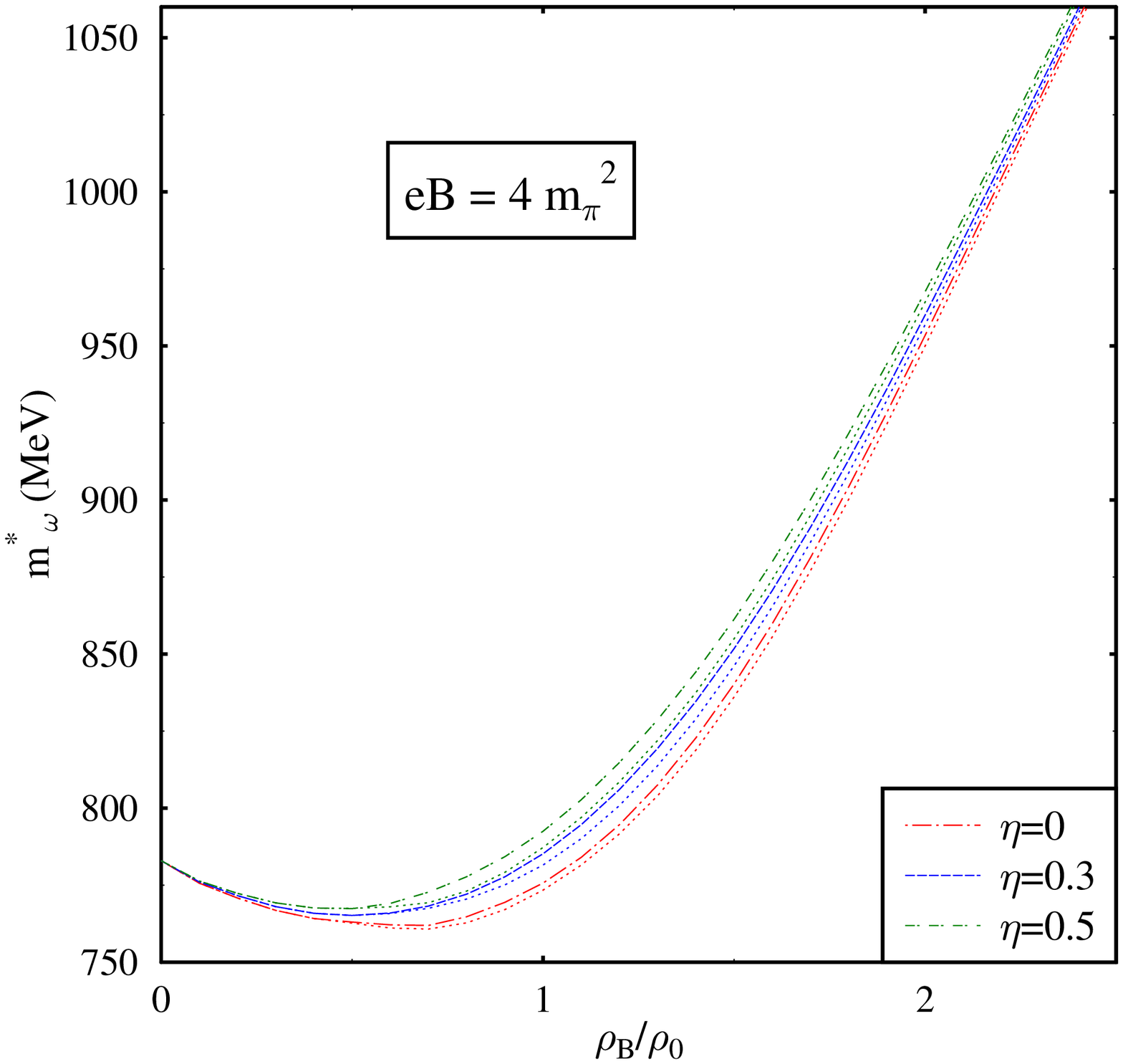}
\caption{(Color online)
The mass of $\omega$ meson plotted as a function of the
baryon density in units of nuclear matter saturation density,
for magnetized nuclear matter
(for $\eta$=0, 0.3, 0.5) with $eB=4 m_\pi^2$.} 
\label{momg_mag_4mpi2}
\end{figure}
\begin{figure}
\includegraphics[width=16cm,height=16cm]{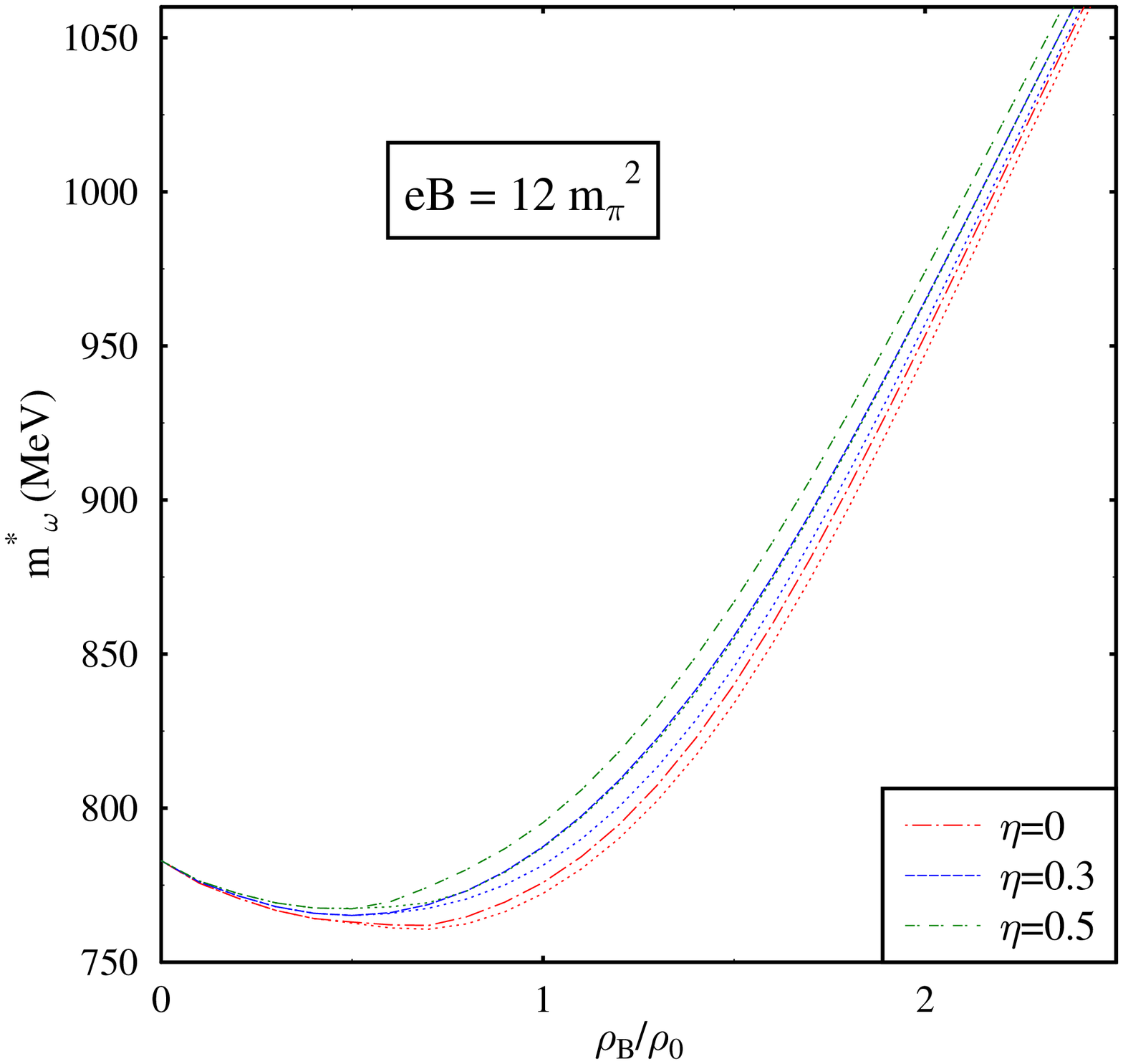}
\caption{(Color online)
The mass of $\omega$ meson plotted as a function of the
baryon density in units of nuclear matter saturation density,
for magnetized nuclear matter
(for $\eta$=0, 0.3, 0.5) with $eB=12 m_\pi^2$.} 
\label{momg_mag_12mpi2}
\end{figure}
\begin{figure}
\includegraphics[width=16cm,height=16cm]{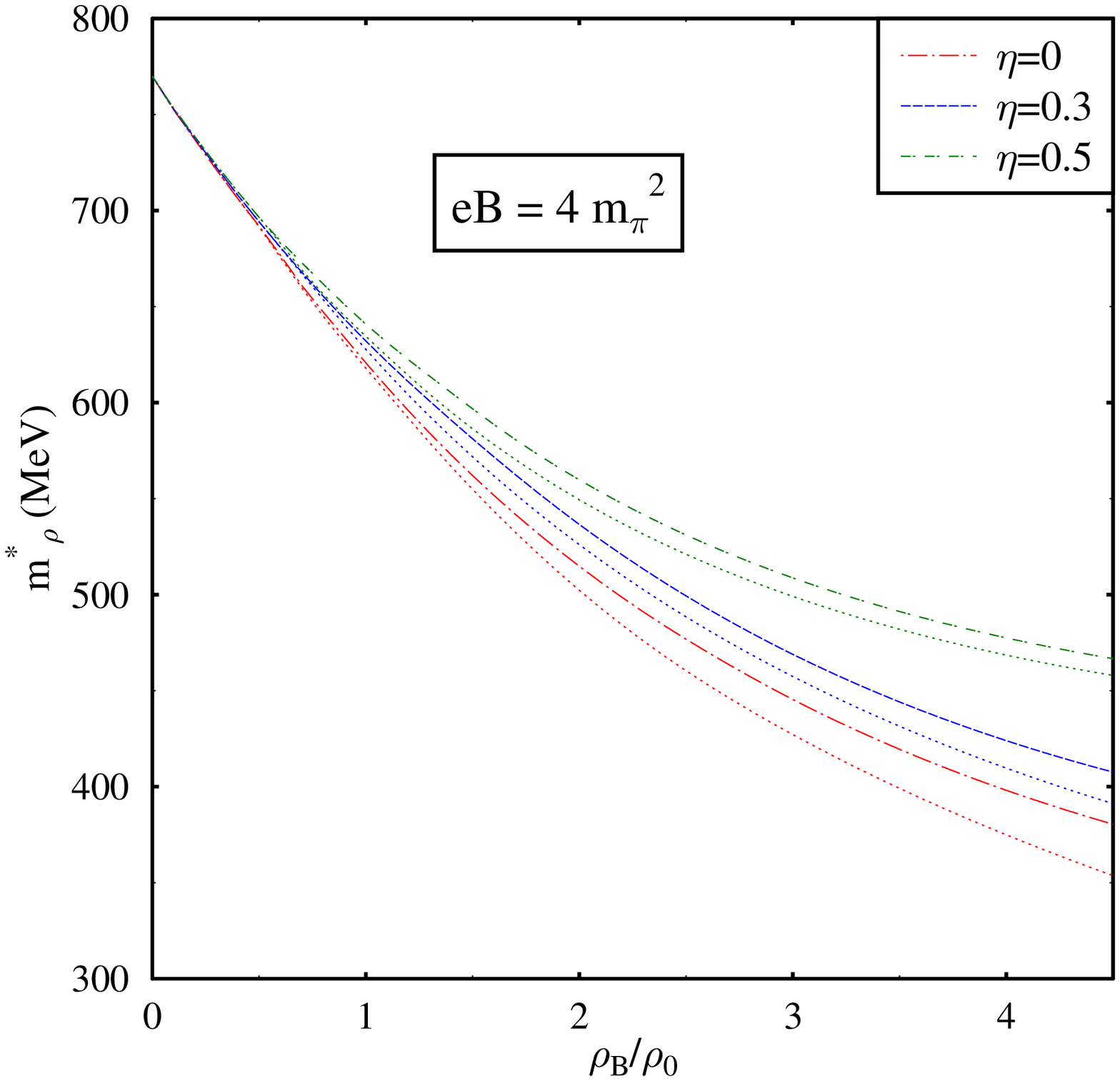}
\caption{(Color online)
The mass of $\rho$ meson plotted as a function of the
baryon density in units of nuclear matter saturation density,
for magnetized nuclear matter
(for $\eta$=0, 0.3, 0.5) with $eB=4 m_\pi^2$.}
\label{mrho_mag_4mpi2}
\end{figure}
\begin{figure}
\includegraphics[width=16cm,height=16cm]{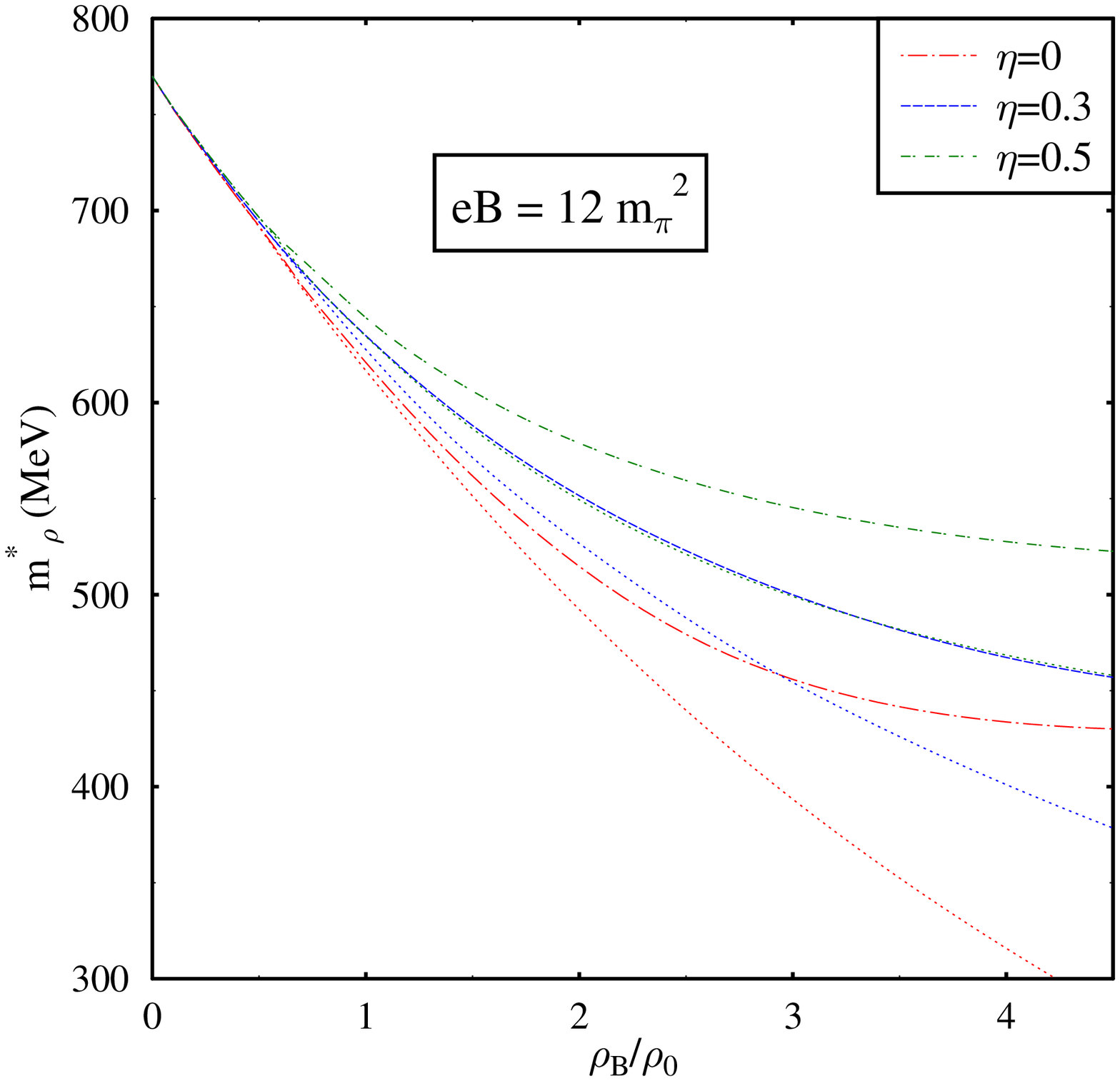}
\caption{(Color online)
The mass of $\rho$ meson plotted as a function of the
baryon density in units of nuclear matter saturation density,
for magnetized nuclear matter
(for $\eta$=0, 0.3, 0.5) with $eB=12 m_\pi^2$.}
\label{mrho_mag_12mpi2}
\end{figure}
\begin{figure}
\includegraphics[width=16cm,height=16cm]{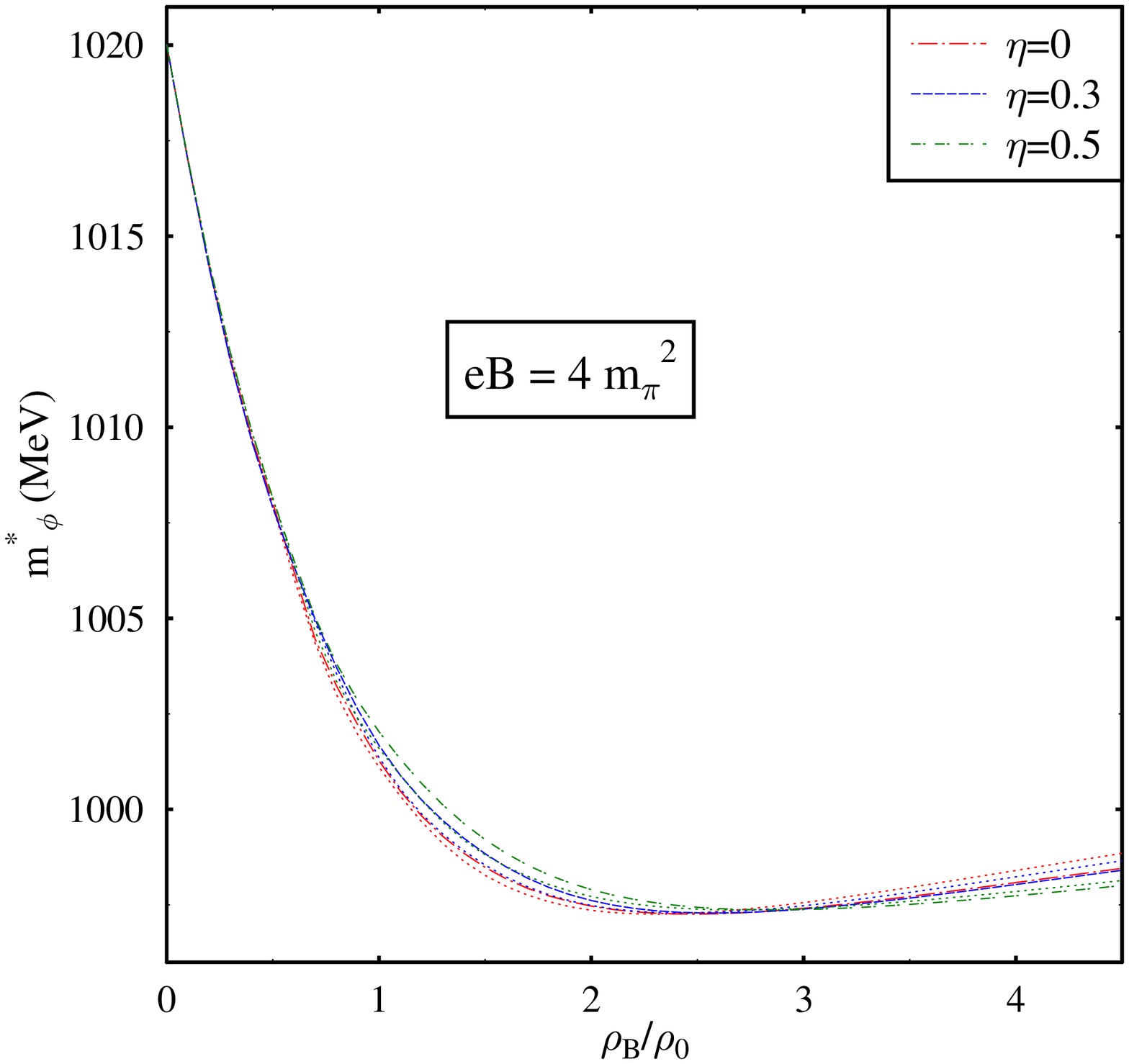}
\caption{(Color online)
The mass of $\phi$ meson plotted as a function of the
baryon density in units of nuclear matter saturation density,
for magnetized nuclear matter
(for $\eta$=0, 0.3, 0.5) with $eB=4 m_\pi^2$.}
\label{mphi_mag_4mpi2}
\end{figure}
\begin{figure}
\includegraphics[width=16cm,height=16cm]{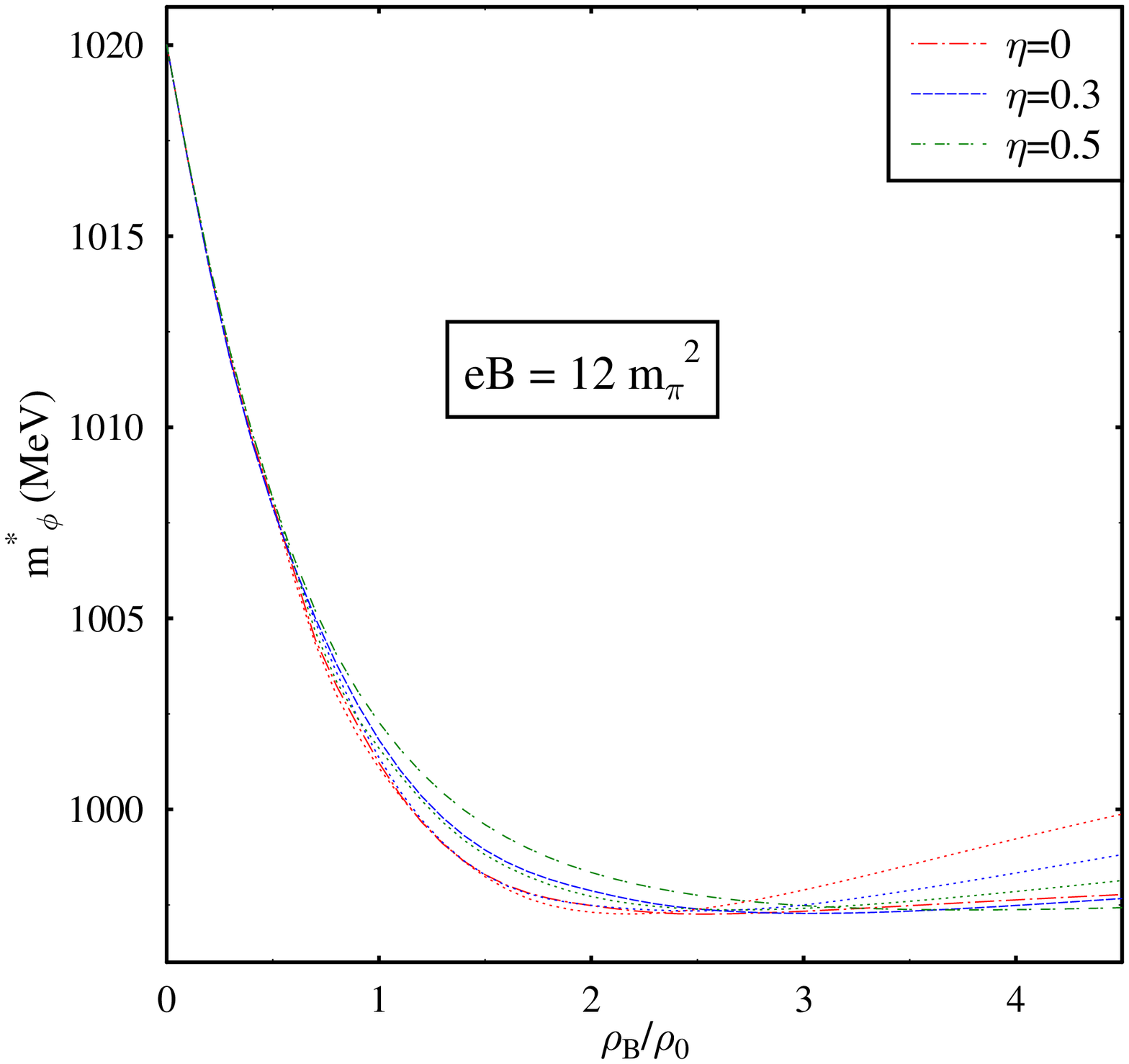}
\caption{(Color online)
The mass of $\phi$ meson plotted as a function of the
baryon density in units of nuclear matter saturation density,
for magnetized nuclear matter
(for $\eta$=0, 0.3, 0.5) with $eB=12 m_\pi^2$.}
\label{mphi_mag_12mpi2}
\end{figure}

\section{Results and Discussions}

In this section, we study the in-medium masses 
of light vector mesons ($\omega$, $\rho$ and $\phi$)
in strongly magnetized nuclear matter, 
using the QCD sum rule approach from the light quark condensates 
and the scalar
gluon condensate calculated within a  chiral SU(3) model. 
The broken scale invariance of QCD is incorporated 
in the effective hadronic model 
through a scale breaking logarithmic potential 
in terms of a scalar dilaton field, $\chi$ and 
the gluon condensate in the magnetized nuclear matter
is calculated from the medium modification 
of this scalar dilaton field. 
In the chiral effective model, the calculations are done 
in the mean field approximation. In mean field approximation, 
the meson fields are treated as classical fields. 
The nonstrange light quark condensates ($\langle \bar u u\rangle$, 
$\langle \bar d d\rangle$) and strange condensate ($\langle \bar s s\rangle$) 
in the asymmetric nuclear matter
in the presence of magnetic field are calculated within the
model from the values of the nonstrange ($\sigma$, $\delta$)
and strange ($\zeta$) scalar fields within the chiral SU(3) 
model. 
For given values of baryon density, $\rho_B$, 
the magnetic field, and, the isospin asymmetry
parameter, $\eta=({\rho_n-\rho_p})/({2\rho_B})$ 
($\rho_p$ and $\rho_n$ are the proton and neutron 
number densities, respectively),
the mean values of the scalar fields, 
$\sigma$, $\zeta$, $\delta$ and $\chi$ are obtained
by solving the coupled equations of motion of these fields.
In the absence of a magnetic field,
the in-medium masses of the light vector mesons
in asymmetric hadronic matter were studied using 
QCD sum rule approach and using the gluon and quark condensates
calculated within the chiral SU(3) model \cite{am_vecqsr}.
In the present work, the medium modifications of the
vector meson masses are studied in the presence of
a strong magnetic field.

In the present investigation, 
the values of the current quark masses are taken as 
 $m_u$=4 MeV, $m_d$=7 MeV and $m_s$=150 MeV. 
The vacuum masses of the $\rho$, $\omega$ and $\phi$ mesons
are taken to be 770, 783 and 1020 MeV.
The coefficients $\kappa_i$ ($i=q,s$) are solved
from the FESRs in vacuum \cite{am_vecqsr}
and are obtained as 7.788, 7.236 and $-$1.21 
for the  $\omega$, $\rho$ and $\phi$ mesons. 
The difference in the values of $\kappa_q$
obtained from solving the FESRs of the 
$\omega$ and $\rho$ mesons are due to the
difference in their vacuum masses.

In the presence of a magnetic field, the proton has 
contributions from the Landau energy levels. The 
anomalous magnetic moments of the nucleons 
are also taken into consideration in studying 
the in-medium masses of light vector mesons 
in nuclear matter in presence of a magnetic field,
using the QCDSR approach. 
The light quark condensates and the scalar gluon condensate, 
are calculated from the equations (\ref{qqbar}) and (\ref{chiglu}), 
which are then used to calculate the values of the coefficients, 
${c^*}_2^V$ and ${c^*}_3^V$ for the vector mesons,
$\rho$, $\omega$ and $\phi$ as given by equations
(\ref{c2rhomg}), (\ref{c3rhomgf}), (\ref{c2phi}), 
and (\ref{c3phif}).
Using the values of these coefficients, 
the in-medium masses of the light vector mesons
are calculated by solving the coupled equations
(\ref{fesr1mf}), (\ref{fesr2mf}) and (\ref{fesr3mf})
involving the in-medium values $F^*_V$, ${s^*}_0^V$
and $m^*_V$.

In figures \ref{momg_mag_4mpi2} and  \ref{momg_mag_12mpi2}, 
the in--medium masses of $\omega$ mesons 
are plotted as functions of baryon density, 
$\rho_B$ (in units of nuclear matter saturation density, 
$\rho_0$) for values of magnetic fields,
$eB=4 m_\pi^2$ and  $eB=12 m_\pi^2$ respectively,
with values of isospin asymmetry parameter, $\eta$ taken to be
0, 0.3 and 0.5. These masses are plotted including the
anomalous magnetic moments (AMM) of the nucleons
and are compared to the case when the AMMs are not
taken into consideration. The masses of the $\omega$ meson
are also given in Table 1, for the isospin symmetric
($\eta$=0) as well as for the extreme isospin asymmetric case
of $\eta$=0.5, for magnetic fields, $eB=4m_\pi^2,\;8m_\pi^2,
\;12m_\pi^2$, (a) without and (b) with 
the anomalous magnetic moments (AMM) of nucleons
taken into account.
The $\omega$ meson mass is observed to have an initial drop 
with increase in density, for sub-nuclear densities.
This is due to the fact that the contribution of the scattering term, 
which is proprotional to baryon density, $\rho_B$, is small
at low densites, and, the medium modification of the 
$\omega$ meson mass is dominated by modificaitons of the
the light quark condensates in the medium, which leads to
a drop in the mass of the $\omega$ meson.
As the density is further increased,
the effects of the scattering of the vector meson $\omega$
from the nucleons become appreciable, leading
to a rise in the $\omega$ meson mass. This
observed behaviour can be understood from the first two 
FESRs. The effective mass squared of the vector meson, $V$,
is obtained as dividing equation (\ref{fesr2mf}) 
by (\ref{fesr1mf}) as
\begin{equation}
{m^*_V}^2=
\frac { \Big (
\frac {({s^*}^V_0)^2 c^V_0}{2}-{c^*}_2^V \Big )}
{({c^V_0} {{s^*}^V_0} +{c^V_1}) -({1}/{d_V})\cdot 12\pi^2 \Pi^V(0)},
\label{meff2v}
\end{equation}
where, as has already been mentioned, $\Pi^ V(0)={\rho_B}/{(4M_N)}$,
for $V=\omega,\rho$ and $\Pi^V(0)$ vanishes for $\phi$ meson
as $\phi$-nucleon coupling is assumed to be zero in the present 
work. The scattering term for the non-strange vector mesons, 
being proportional to the density,
becomes appreciable at higher densities. This leads to a smaller
value for the denominatior and hence to an
increase in the mass of $\omega$ meson at high densities.
This behaviour of the $\omega$ mass with density
was also observed for the case of zero magnetic field
in Ref. \cite{am_vecqsr}.
The effects of the magnetic field, which are through
the light quark and gluon condensates, are observed
to be much smaller as compared to the density effects.
The values of the $\omega$ meson mass (in MeV) 
at densities $\rho_0$(2$\rho_0$), 
for the values of the isospin asymmetric parameter,
$\eta$=0, 0.3 and 0.5, 
are observed to be 773.4 (950), 781.55 (956.86) and 787.24 (964.2) 
at the value of the magnetic
field as $eB = 4m_\pi^2$, when the anomalous magnetic moments
(AMM) of nucleons are not considered, and, 
775.7 (953.4), 785.2 (960) and 792.5 (967.56),
when AMMs are taken into consideration.
The isospin asymmetry effects are observed to be
larger for the higher value of the magnetic field,
$eB=12 m_\pi^2$, plotted in figure \ref{momg_mag_12mpi2}. 
The contribution of the scattering term which 
is the dominant contribuion to the mass
of $\omega$ meson at high densities,
being proportional to $\rho_B$, is independent
of the isospin asymmetry. This leads to lessening 
of the isospin asymmetry effects at higher densities,
as is evident from the figures \ref{momg_mag_4mpi2}
and \ref{momg_mag_12mpi2}. 
The isospin asymmetry effects on the $\omega$ mesons mass
are observed only at densities, of around 0.5$\rho_0$ to 
about 2 $\rho_0$, above which these effects are seen to be
diminished as the scattering effects start becoming important.
The isospin asymmetry effects are observed to be more
appreciable for the higher value of the magnetic field,
$eB=12 m_\pi^2$, as can be seen from figure \ref{momg_mag_12mpi2}. 

The density, isospin asymmetry effects on the masses 
of the $\rho$ meson are  illustrated in figures \ref{mrho_mag_4mpi2}
and  \ref{mrho_mag_12mpi2} for the values of $eB$ as $4m_\pi^2$
and $12 m_\pi^2$ respectively. The in-medium $\rho$ meson
masses are given in Table 2, for typical values of the magnetic field,
density and isospin asymmetry parameters, which are calculated
(a) without and (b) with the anomalous magnetic moments (AMM) 
of the nucleons.
The $\rho$ meson mass in the presence of a magnetic field
is observed to drop with baryon density,
similar to the case of zero magnetic field studied 
previously \cite{am_vecqsr}.
The contribution of the scattering of the $\rho$ meson
due to the nucleons in the nuclear matter is small,
due to the factor $(1/{d_V})$ in this term,
which makes the contribution of the Landau scattering term 
to be 9 times smaller than that of the $\omega$ meson,
as ${(1/d_\rho)}/{(1/d_\omega)}=1/9$. 
The effects of the isospin asymmetry
are observed to be large at high densities,
as was seen for the case of zero magnetic field
\cite{am_vecqsr}. The effects of magnetic fields
as well as of anomalous magnetic 
moments are observed to be appreciable at high
densities, as can be seen in Table 2.
Accounting for the anomalous magnetic moments (AMM)
of the nucleons, the values of the $\rho$ meson mass 
for symmetric nuclear matter
($\eta$=0) at densities $\rho_0$(3$\rho_0$), 
are 620.7 (445.4) and 620.9 (455.9) for $eB = 4m_\pi^2$,
and $eB = 12m_\pi^2$ respectively.
These values are modified to 632.1 (468.9)
and 634.85 (500) at $\eta=0.3$ and 
640.85 (508.7) and 644.2 (545.36)
at $\eta$=0.5, for the same magnetic fields.
The drop in the $\rho$ meson mass is observed to be
smaller when the AMMs of the nucleons are taken
into consideration. The isospin asymmetry effect
is observed to be much larger for the larger magnetic
field, $eB=12 m_\pi^2$, especially at higher densities,
as can be seen in figure \ref{mrho_mag_12mpi2}. 
In Table 2, the values of the $\rho$ meson mass in the presence
of magnetic field for the symmetric nuclear matter as
well as for asymmetric nuclear matter with $\eta$=0.5,
at densities $\rho_0$ and $4\rho_0$, are compared with the values
obtained for zero magnetic field in Ref. \cite{am_vecqsr}.
As can be seen from Table 2, in the absence of a magnetic field, 
the in-medium $\rho$ mass calculated at the nuclear matter saturation 
density, $\rho_0$, in symmetric nuclear matter,
is 622.2 MeV \cite{am_vecqsr}, 
which is similar to the value obtained  
in Ref. \cite{hatlee}, using the linear density approximation. 
The mass of $\rho$ meson at $\rho_0$ at zero magnetic field,
may be compared with the value of 670 MeV
obtained using QCD sum rule approach in Ref. \cite{kwonprc2008},
and, of around 530 MeV, in an improved QCD sum rule calculation 
\cite{asakawako}. In the present work, the effects of magnetic field
on the $\rho$ meson mass in isospin asymmetric nuclear matter
have been investigated.

The in-medium masses of the $\phi$ meson, plotted in figures 
\ref{mphi_mag_4mpi2} and \ref{mphi_mag_12mpi2}
for the values of $eB$ as $4m_\pi^2$ and $12 m_\pi^2$ 
respectively,
are observed to decrease sharply with density upto 
a density of around $\rho_0$, above which there is
observed to be very less modification in the mass
of the $\phi$ meson.
There is no contribution to the $\phi$ mass
from the scattering term.
This is due to the fact that the nucleon--$\phi$ meson coupling 
is zero in the parameter set chosen for the vector meson-baryon 
interactions within the chiral SU(3) model.
It might be noted here that $\phi \rightarrow K\bar K$
is OZI allowed, whereas OZI rule forbids the decay
of $\phi$ meson to pions. There is, however, observed to be 
violation of OZI rule by about 5$\%$
in the $\phi \rightarrow 3\pi$ channel \cite{klinglnpa,klinglzpa}.   
The decay $\phi \rightarrow 3 \pi$ and the decay
of $\omega \rightarrow 3 \pi$ (through direct decay
as well as the two step decay, $\omega \rightarrow \rho \pi$ 
followed by $\rho \rightarrow 2 \pi$) imply that there 
is mixing between the $\phi$ meson
and the non-strange ($\rho$, $\omega$) vector mesons.
However, the decay of $\phi$ is dominated by the OZI allowed
$\phi \rightarrow K \bar K$ and the mass modification
of the $\phi$ meson is predominantly due to the change
of the strange condensate in the hadronic medium. 
The drop in $\phi$ mass is much smaller than
the mass drop of the $\rho$ meson (where the
scattering term has negligible contribution)
due to the large drop of the nonstrange quark
condensate as compared to the drop
of the strange quark condensate in the nuclear medium.
At higher densities, the strange condensate
remains almost constant and hence the $\phi$ meson
mass changes very little with density,
whereas the nonstrange quark condensate
continues to drop as density is increased,
leading to a monotonic drop of the 
$\rho$ meson mass with density.
These behaviours of the $\rho$ and $\phi$
vector meson masses
were also observed for the case of zero magnetic field
\cite{am_vecqsr}.
The effects of anomalous magnetic moments of
the nucleons are observed to be larger at higher
densities, even though the magnitudes of these
still remain small. The strange quark 
condensate  as well as the scalar gluon condensate have very small
effects from isospin asymmetry, leading to the modifications
of the $\phi$ meson mass to be very similar in the isospin symmetric
and asymmetric nuclear matter.

\section{Summary}
In summary, in the present investigation, we have calculated 
the masses of the light vector mesons ($\omega$, $\rho$ and
$\phi$) in the nuclear matter in the presence of a strong 
magnetic field,
using QCD sum rule approach, due to the modifications 
of the light quark and scalar gluon condensates, calculated 
within a chiral effective model. The medium modifications 
are due to the changes of the light quark condensates and 
the scalar gluon condensate in the magnetized isospin 
asymmetric hadronic matter, which are calculated within 
mean field approximation, from the changes in the expectation 
values of the scalar fields ($\sigma$, $\zeta$ and $\delta$) 
and a dilaton field ($\chi$) 
from their vacuum values, respectively, within a chiral SU(3) 
model. The masses of the $\rho$  mesons are dominantly 
governed by the nonstrange quark condensates which
lead to decrease in these masses with increase in density.
For $\omega$ meson, the drop in the nonstrange
quark condensates leads to a drop in the mass 
at sub-nuclear matter densities, which however, is
overcome by the  $\omega$-nucleon scattering term,
leading to an increase in the $\omega$ mass 
as the density is further increased.
The scattering term which dominates for the 
$\omega$ meson mass at high densities,
is of the form $(\rho_p+\rho_n)$ and is 
thus independent of the isospin asymmetry 
of the nuclear medium.
The mass of $\phi$ meson is dominated by the behaviour
of the strange condensate, $\langle \bar s s \rangle$
in the medium, leading to an initial
drop with density and the mass changes slightly
as the density is further increased.
The isospin asymmetry effects are observed 
to be more for $\rho$ meson as compared
to the $\omega$ and $\phi$ mesons. 
The effects of anomalous magnetic moments of the nucleons 
are seen to be appreciable at higher densities and higher 
magnetic fields for the $\rho$ meson mass. The 
density effects are observed to be the dominant medium 
effects on the masses of these light vector mesons.

\acknowledgements
One of the authors (AM) is grateful to ITP, University of Frankfurt,
for warm hospitality and acknowledges financial support from Alexander
von Humboldt foundation when this work was initiated.

\end{document}